# Detection of a possible superluminous supernova in the epoch of reionization

Jeremy Mould*, Tim Abbott[†], Jeff Cooke*, Chris Curtin*, Antonios Katsianis[¶], Anton Koekemoer[Δ], Edoardo Tescari[♦], Syed Uddin[#], Lifan Wang[✪], Stuart Wyithe[♦]

**Short communication**

The Epoch of Reionization (EoR) poses a number of puzzles, including the sufficiency of ionizing radiation, the rapid production of supermassive black holes and the early appearance of abundant metals in larger galaxies. For the first of these, solutions focus on the main sequence evolution of massive stars, but the second and third puzzles involve the terminal phase of stellar evolution, and in particular, core collapse supernovae.

Candidates for the ionizing galaxies are seen in deep surveys with the Hubble Space Telescope and wide surveys with large telescopes, such as Subaru. Their supernovae would also be expected to be visible, especially superluminous supernovae (SLSN), such as those seen[1,2] at lower redshifts. Here we present a candidate for the first EoR SLSN, discovered in our deep field survey with the Dark Energy Camera at the Blanco telescope of the Cerro Tololo Interamerican Observatory. The DECamERON survey[3] has accumulated deep near infrared images (0.7—1.05 microns) in three circumpolar fields over the lifetime of the camera, which together provide a solid reference for finding transient objects.

One of these fields[4], New Southern Field-1 (NSF-1) was observed on four nights in 1'' seeing in August 2016. Pipeline processing by NOAO caused individual 5 minute exposures to stacked. After downloading the resultant images from NOAO, we carried out further stacking. The SUDSS program[4] contributed to the reference images. Stellar photometry with a 0.8'' aperture was carried out on the images with DAOPHOT, and cuts were applied to single out objects that had (1) brightened by a magnitude from the reference images and that (2) had a flux deficit of at least a magnitude in the *i* filter relative to the *z* filter in the reference image. The second cut is motivated by the dropout concept that a host galaxy with redshift 6 presents its Lyman break to the *i* filter. Figure 1 shows our top candidate and Figure 2 its spectral energy distribution (sed), using standard calibrations https://cdcvs.fnal.gov/redmine/projects/des-sci-verification/wiki/Photometry.

*Centre for Astrophysical & Supercomputing, Swinburne University, Hawthorn 3122, Australia

[†]Cerro Tololo Interamerican Observatory, NOAO, La Serena, Chile

[Δ]Space Telescope Science Institute, Baltimore, MD 21218, USA

[#]Purple Mountain Observatory, Chinese Academy of Sciences, Nanjing, Jiangsu Sheng, China

[✪]Texas A&M University, College Station, TX 77843, USA

[♦]University of Melbourne, School of Physics, Melbourne 3010, Australia

[¶]Departamento de Astronomia, Universidad de Chile, Las Condes, Santiago, Chile

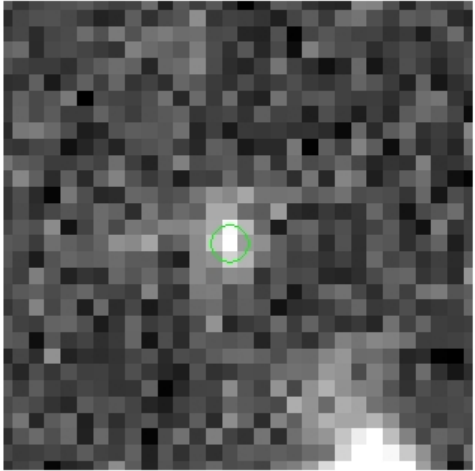 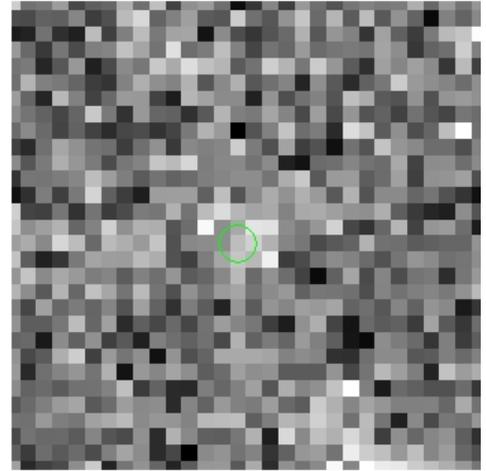
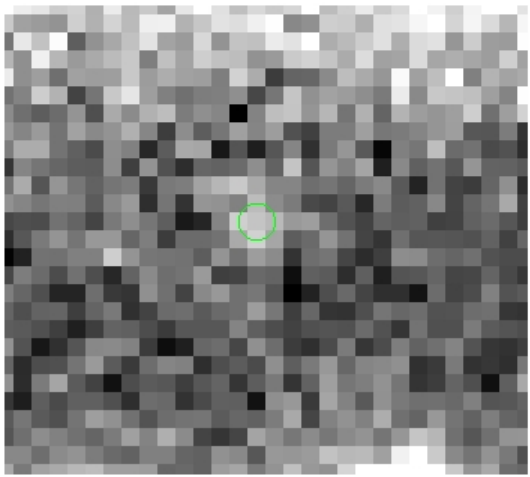 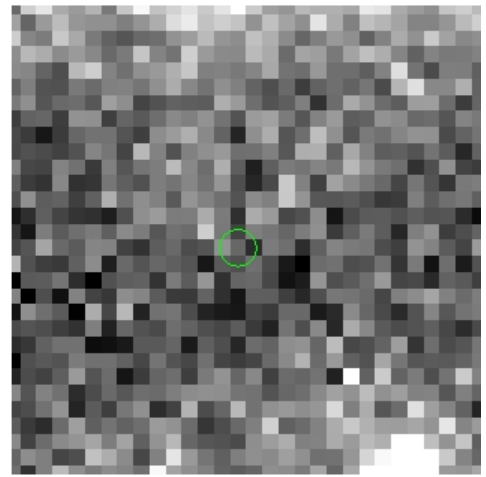

Figure 1 – The upper two images are supernova discovery images in *z* and *i*; the lower two are the prediscovery images obtained from the preceding three years. The DECam scale is 0.27'' per pixel, and the green 0.55'' aperture is the astrometric position of the transient. Stack time is shown in bold at the top left of individual 'postage stamps'. We take the host to be the visible galaxy at the centre of the frame, ruling out the bright object at the lower right of the frame, which is evidently not an *i* band dropout. Only the top pixels of that object are seen, and its separation from the SLSN candidate is 23 kpc at z=6. We have not pursued the astrometry of DECam pipeline images beyond integral pixel precision, although DES has verified their astrometry to better than 100 mas.

Alternative transient hypotheses seem unlikely: observations obtained at 10 epochs 2013-2015 show no intrinsic variations inconsistent with photon statistics, suggesting this is not an optically variable AGN, and the *z*-Y colour is bluer than a Galactic halo flare star. *Kepler* data[5] show that flare stars on the main sequence are cooler in effective temperature than 4000K and two thirds of them are cooler than 3000K. Had we mistaken the 400 nm Balmer break at redshift 2 for the 120 nm Lyman break, the absolute *z* magnitude would still be 2 mag brighter than a SNIa. A more normal SNIa (similar to SN Primo[6]) at z = 1.3 would have *r*~27 mag, *i*~26 mag and *z*=Y~25.0 mag. That is redder in *i*-*z* than J211451.47-654102.5. The possibility of a tidal distortion event[7] cannot be dismissed. The absence of resolved galaxies in the field lends no encouragement to the possibility of a lensed lower redshift supernova.

The AB magnitudes of J211451.47-654102.5 are 25.0 +/- 0.14 in the Y band and 24.97 +/- 0.08 in z. The brightest SLSN candidate to date[15] is ASASSN-15lh at a redshift of 0.23. ASASSN-15lh at z = 6 would have a magnitude at maximum light that is close[1] to this value. Figure 3 shows its UVOT sed fitted to a blackbody[1] and redshifted to z = 6 (lower plot). Like Figure 2, the sed is fairly flat across the *iz*Y bands, as these bands are not far from the peak of the sed. Integrating the energy distribution of ASASSN-15lh, we estimate J211451.47-654102.5 has a luminosity of 2 x $10^{11}$ $L_\odot$ assuming z = 6 and a standard cosmology. ASASSN-15lh itself may be a tidal disruption event[8]. Less ambiguous SLSNe are PTF and Pan-STARRS transients[9] with $AB_U$ ~ -22 mag.

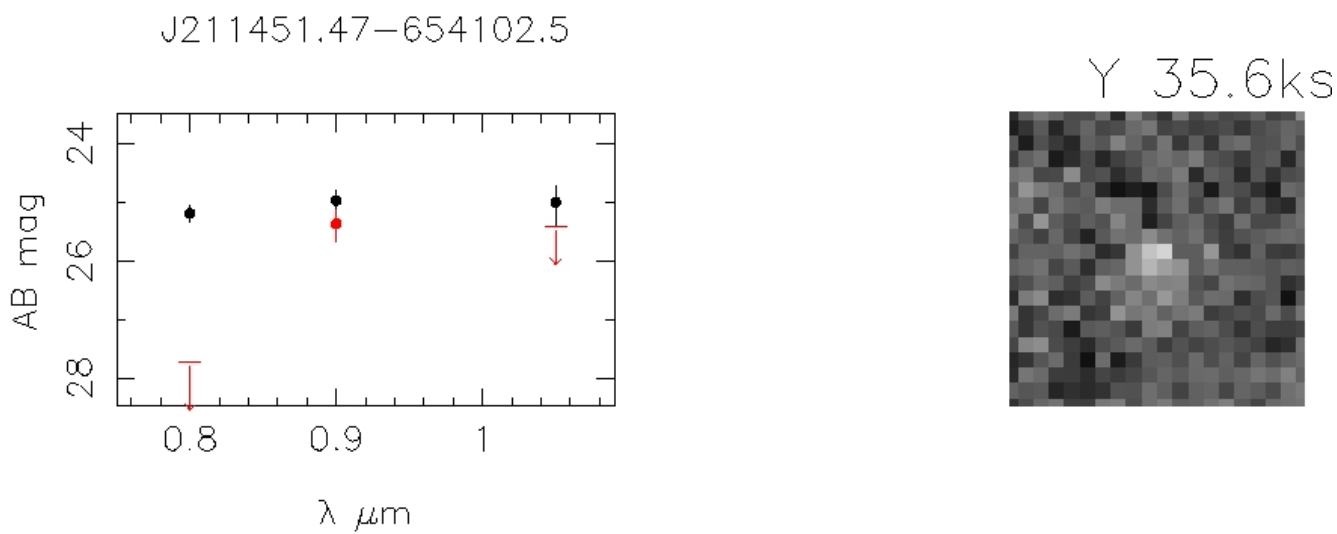

Figure 2 – (left) The spectral energy distribution of J211451.47-654102.5. Black points are August 2016; red points are archival. The error bars are one sigma. On the right is the stacked Y band image.

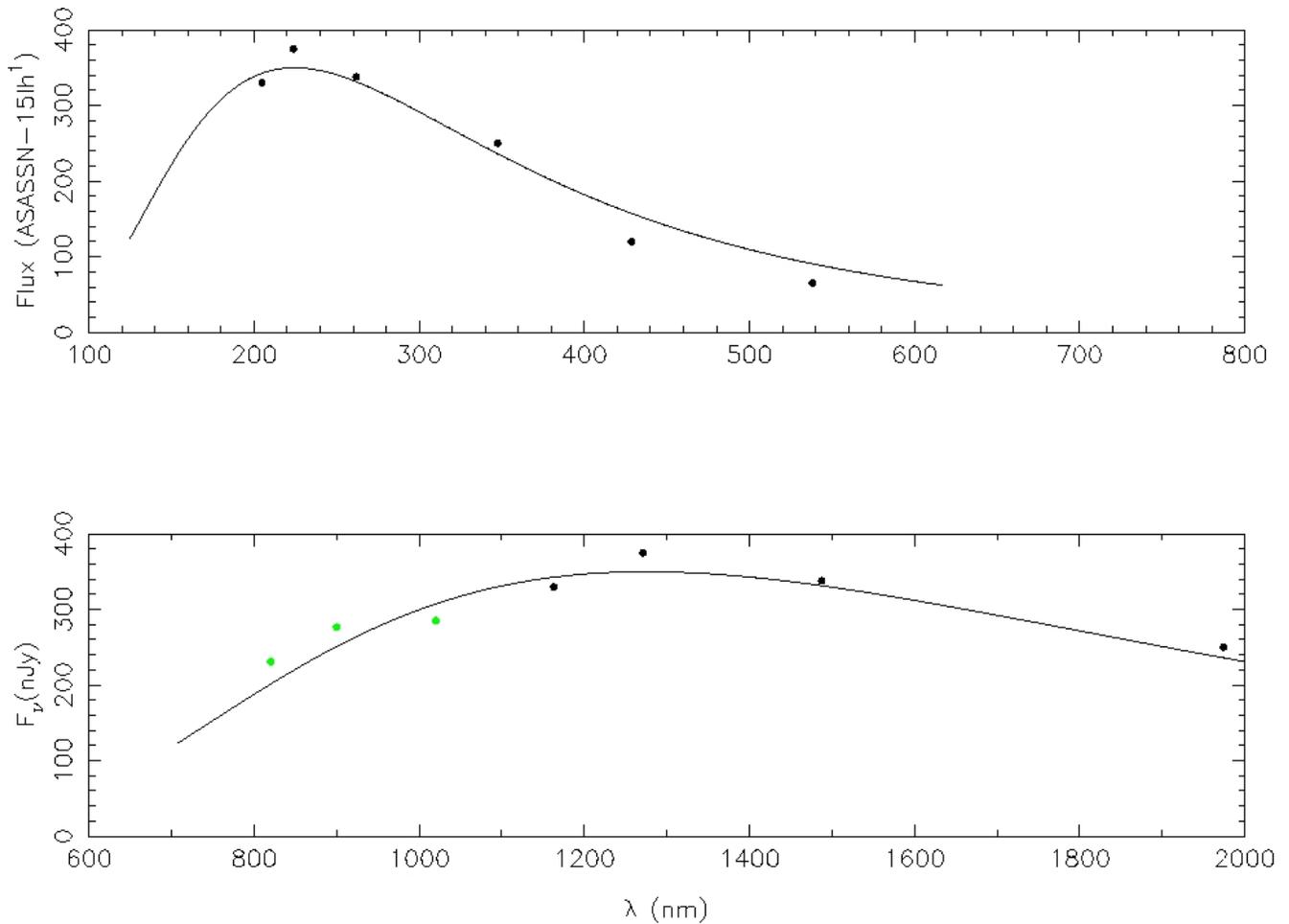

Figure 3 – The spectral energy distribution of ASASSN-15lh from first epoch UVOT data[1]. The curve is a redshifted blackbody of temperature 15860K. The lower curve is the same object and fit redshifted to z = 6. The green points are the fluxes of J211451.47-654102.5.

The dominant uncertainty in the photometry at all three wavelengths is sky background. We have ascertained this uncertainty by inserting artificial stars into the stacked data. In this way we can be sure that the one-sigma upper limit on the *i* magnitude of the supernova's location before our 2016 epoch is $AB_i$ = 27.8 mag. The two-sigma upper limit is 26.5 mag. The data are recorded in Table 1.

|  | *i* | *z* | Y |
| --- | --- | --- | --- |
| 2013-2015 | 27.5 | 25.37 | 25.5 |
| uncertainty | +1 | +0.29 -0.23 | +0.4 -0.1 |
| 2016 | 25.19 | 24.97 | 25.0 |
| uncertainty | +0.16 -0.14 | +0.19 -0.16 | +0.38 -0.28 |

We noted earlier that no previous variation in the archival data had been detected beyond what might be expected from photon statistics. In Figure 4 we show the individual epochs of stacked *z*- and *i*-band data. The explosion of the SLSN is indicated by these data to have occurred after 1/1/2016. A SLSN that remains within a magnitude of the peak luminosity for 100 days in the rest frame would do so for 2 years, when observed at z = 6. The simplest view of the light curve is that the host galaxy had *z* = 25.7 mag and *i* > 27 mag until at after 1/1/2016. But we cannot rule out the possibility that a brightening from z = 26 mag to z = 25.5 occurred as early as 1/1/2015.

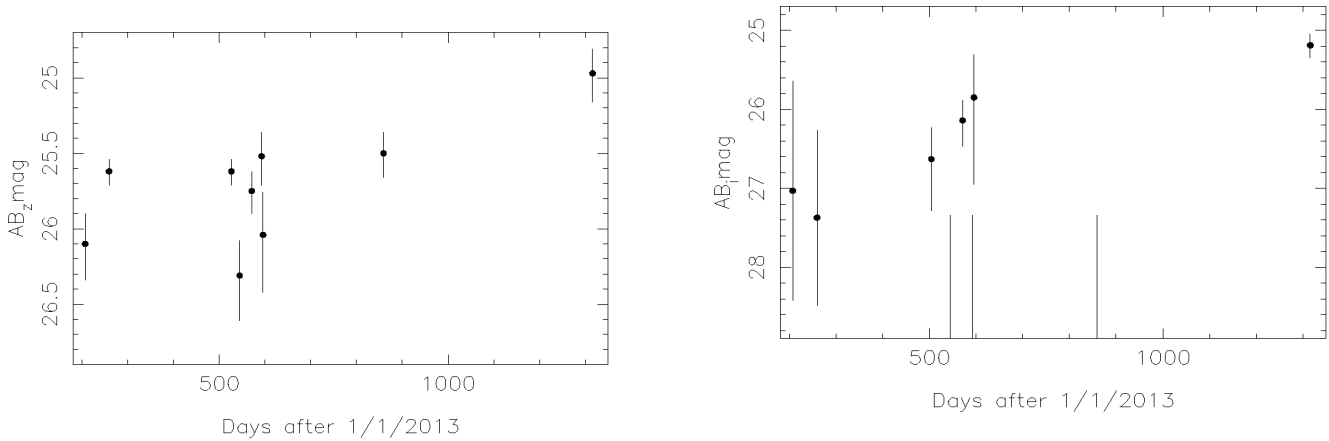

Figure 4 - Archival photometry of transient J211451.47-654102.5 in *z* and *i* bands.

For completeness we also show archival non-detections in *g* and *r* in Figure 5.

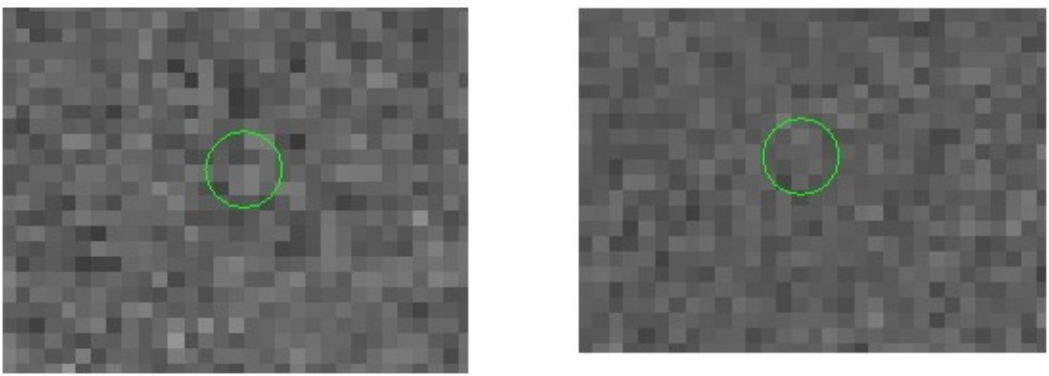

Figure 5 - Archival *g* band (left) and *r* band (right). The superposed circles have a diameter of 1.2''.

Mapping the neutral hydrogen at high z and finding the UV emitters at this epoch are both vital approaches to understanding the EoR. Assuming that J211451.47-654102.5 is the first of many z = 6 SLSNe, we have a third probe of the linkage between the intergalactic medium and star formation, an approach that opens many questions about the first few hundred Myrs.

Four key questions are the initial mass function (IMF) of high z SLSNe, the physical model, their frequency of occurrence and their energy output. A top heavy IMF is supposed in most models, but undemonstrated observationally. Physical models of SLSNe include pair instability supernovae (PISNe)[10], magnetars[11], quark novae[12], radiatively shocked circumstellar matter[13], and jet-cocoon structures[14]. The frequency of SLSNe can be estimated as ~6/yr/sq deg and the energy output may be as high as $10^{55-56}$ ergs, exceeding the main sequence radiated energy of 100 $M_\odot$ stars at $10^{54}$ ergs.

To explore these questions, we need to enlarge our surveys, as finding one candidate is not a rate measurement or a luminosity distribution. We need a redshift for J211451.47-654102.5 or its host. In cycle 25 we are applying for Hubble Space Telescope observations of J211451.47-654102.5 and other similar DECam transients. We need progress with physical models of SLSNe. The future looks bright for the study of superluminous supernovae at redshifts from that of Population III onwards.

In summary, an interesting transient has been detected in one of our three Dark Energy Camera deep fields. Observations of these deep fields take advantage of the high red sensitivity of DECam, a Cerro Tololo Interamerican Observatory 4m telescope instrumental development which the principal investigator has been following to fruition for over a decade. The survey includes the Y band with rest wavelength 1430Å at z = 6. Survey fields (the Prime field 0555-6130, the 16hr field 1600-75 and the SUDSS New Southern Field) are deeper in Y than the VISTA and related infrared surveys. They are circumpolar, allowing all night to be used efficiently, exploiting the moon tolerance of 1 micron observations to minimize conflict with the Dark Energy Survey. As an *i*-band dropout (meaning that the flux decrement shortward of Lyman alpha is in the *i* bandpass), the transient we report here is a supernova candidate z ~ 6, with a luminosity comparable to the brightest known current epoch superluminous supernova (i.e., ~ 2 x $10^{11}$ $L_\odot$).

**Acknowledgements** The data presented herein were obtained at NOAO's Cerro Tololo Interamerican Observatory, which is operated by AURA for the U.S. National Science Foundation.

This project used data obtained with the Dark Energy Camera (DECam), which was constructed by the Dark Energy Survey (DES) collaboration. Funding for the DES Projects has been provided by the DOE and NSF (USA), MISE (Spain), STFC (UK), HEFCE (UK). NCSA (UIUC), KICP (U. Chicago), CCAPP (Ohio State), MIFPA (Texas A&M), CNPQ, FAPERJ, FINEP (Brazil), MINECO (Spain), DFG (Germany) and the collaborating institutions in the Dark Energy Survey, which are Argonne Lab, UC Santa Cruz, University of Cambridge, CIEMAT-Madrid, University of Chicago, University College London, DES-Brazil Consortium, University of Edinburgh, ETH Zurich, Fermilab, University of Illinois, ICE (IEEC-CSIC), IFAE Barcelona, LBL, LMU Munchen and the associated Excellence Cluster Universe, University of Michigan, NOAO, University of Nottingham, Ohio State University, University of Pennsylvania, University of Portsmouth, SLAC National Lab, Stanford University, University of Sussex, and Texas A&M University.

**Author Contributions** J.M. is the principal investigator of the DECamERON project. Other team members performed the observations and contributed to the paper. We thank an anonymous referee for comments which improved the paper.